\documentclass[aip,reprint,pra,floatfix,showpacs,amsmath,amssymb,onecolumn,longbibliography]{revtex4-2}
\usepackage{comment}
\usepackage{subfigure}
\usepackage{float}
\usepackage{siunitx}
\usepackage[utf8]{inputenc}
\usepackage[T1]{fontenc}
\usepackage{mathptmx}
\usepackage{etoolbox}
\usepackage{graphicx}
\usepackage{dcolumn}
\usepackage{bm}

\usepackage[utf8]{inputenc}
\usepackage[T1]{fontenc}
\usepackage{mathptmx}
\usepackage{etoolbox}
\usepackage{graphicx}  
\usepackage{dcolumn}   
\usepackage{bm}        
\usepackage{amssymb}
\usepackage{ amsmath,amstext,accents}   
\usepackage{siunitx}
\usepackage{longtable}
\usepackage{ntheorem}
\usepackage{xcolor}
\usepackage{ulem}
\usepackage{float}
\usepackage{placeins}
\usepackage{datetime}
\usepackage{longtable}
\newcommand{\angstrom}{\textup{\AA}}
\renewcommand{\v}[1]{\mathbf{#1}}

\def\amor{amorphous }

\def\ab{{\it ab-initio }}

\def\twop{$1s2p(^1P)$}

\def\k{\v{k}}
\def\k0{\v{ k}_0}

\def\colmang {\angstrom}

\newcommand{\quotes}[1]{\lq\lq #1\rq\rq}

\def\etal{\textit{et al.}}

\newcommand{\Wh}{ Colm T. Whelan }
\newcommand{\sr}[1]{_{\text{#1}}}

\def\keV{\mathrm{keV}}

\def\odu{Physics Department, Old Dominion University, Norfolk,
Virginia, USA}
\def\camchem{Yusuf Hamied Department of Chemistry, University of Cambridge, Lensfield Road,
Cambridge CB2 1EW, UK}

\def\ie{i.e. }

\def\threeo2{\frac{3}{2}}

\def\af{Asunci{\'o}n Fern{\'a}ndez}
\def\bl{Bertrand Lacroix}

\def\dena{atoms/$\colmang^3$}

\renewcommand{\v}[1]{\mathbf{#1}}

\def\odu{Physics Department, Old Dominion University, Norfolk,
Virginia,23529, USA}

\def\dena{ atoms  $\,\colmang{}^{-3}$}

\newcommand{\CTW}[1]{{\color[HTML]{1624DC}#1}}

\begin{document}
\begin{flushleft}
\end{flushleft}

\title{ \CTW{On the characteristics of helium filled  nano-pores   in \amor  silicon thin films}}
\author{{\bl}}
\affiliation{Departamento de F\'\i sica Aplicada I, Escuela Polit\'ecnica Superior, Universidad de Sevilla, Virgen de \'Africa 7, 41011 Sevilla, Spain}
\author{\af}
\affiliation{Instituto de Ciencia de Materiales de Sevilla, CSIC-Univ. Sevilla, Avda. Am{\'e}rico Vespucio 49, E-41092 Sevilla, Spain}
\author{{N. C. Pyper}}
\affiliation{\camchem}
\author{{Alex J. W. Thom}}
\affiliation{\camchem}
\author{{\Wh}}
\email{cwhelan@odu.edu}
\affiliation{\odu}
\begin{abstract}
{The properties of helium-filled nanopores in amorphous silicon are elucidated by combining theoretical knowledge of helium electronic structure with the results of Scanning Transmission electron microscopy/electron energy loss spectroscopy (STEM/EELS). }
Two of the properties determined are the density and pressure of the confined helium, these being key properties which are needed for application. 
{The experimental data consists, firstly of the shift of the helium $1s^2\rightarrow1s2p(^1P)$ excitation energy from that of a free atom upon entering a condensed phase, and secondly, the intensities of both the elastically and inelastically scattered electron beams.
Analysis uniting theory with the STEM/EELS measurements for both the helium-filled pores and earlier similar studies of helium encapsulated as bubbles in solid silicon
is combined with fully trustworthy and closely related data for helium in its bulk condensed phases.
 The non-empirical theory used, containing no free parameters, is validated by the excellent agreement between the energy shifts predicted for bulk condensed helium with the entirely independently measured values. 
The above comparisons between the pores and bubbles with the bulk material show that the helium behaviour is essentially the same in all three of these environments.
This means that, although the pressure is clearly temperature dependent, the other important properties are governed solely by the helium density. 
 Although the energy of the electron beam used for the pore measurements differs from that in the bubble experiments, it is shown that scaling the results of either the pore or bubbles results using standard scattering theory almost exactly reproduces the experimental results for the other system. This provides confirmatory evidence that the behaviour of the helium in the pores is essentially the same as that in the bubbles.
Further applications require knowledge of the volumes of the pores, which are derived from accurate pore depths.
 A novel method for deducing these depths is presented and shown to introduce smaller experimental errors than the standard approach.

}
\end{abstract}
\maketitle
\section{Introduction}\label{sec1}

The study  of  the properties of individual inert gas atoms, especially helium, trapped under intense pressures in  different materials  is not only of intrinsic interest but is also of wide technological importance. In materials irradiated with energetic helium, nanometer-sized helium bubbles are formed with additional evolution under neutron bombardment\cite{jian24}. Such bubbles are of particular interest for the nuclear industry since they impact
 nuclear waste disposal \cite{malkovsky20} and  readily accrue in the metallic cladding of nuclear reactors \cite{zinkle13} which
 cause the  materials to swell and   become brittle and fracture\cite{zinkle13}. In  prototype fusion devices, nano-bubbles have been found to degrade reactor surfaces by converting them into hair-like materials (e.g. \quotes{tungsten fuzz}) causing erosion of surface atoms into the plasma and quenching of the fusion-reaction\cite{hammond20,reinhart22}.
A second area  of industrial significance is in the nano-structuring of thin films and surfaces obtained by low-energy He plasma treatments and He incorporation via magnetron sputtering
\cite{godinho13,godinho14,shi05}. 

The damage to reactor walls caused by the formation of bubbles of helium\cite{takamura06,baldwin08,tokitani11,zinkle13,hammond20,reinhart22} was the motivation for
most, if not all, of the many studies of such bubbles\cite{donnelly80,jager82,walsh00,taverna08,frechard09,david11,david14,alix15,alix18,sugar18,taylor22} having been focused on determining the pressure exerted on the host material. For thin film applications one needs to move beyond the purely  barometric and also  consider other important characteristics. For studies of helium nano-pores encapsulated in thin films, knowledge of the volume and density of helium in the pores is important for optical enhancement\cite{godinho13,hernadez15}, for the manufacture of solid targets used for nuclear physics measurements\cite{fernadez20} and also for studies involving the catalytic combustion of hydrogen \cite{giarratano18}.
For applications as electrodes for Li-ion batteries (showing  better structural stability during Li charge and discharge cycling) the nano-porosity (volume occupied by the pores) is the most significant parameter \cite{sakabe18}.
In this paper we present a systematic study, using the insights, both theoretical and experimental, gained from earlier studies of helium in bubbles, to illuminate the properties of the helium aggregates in nano-pores of \amor silicon thin films.

Direct measurement of the properties of helium aggregates  encapsulated in a host material is not technically possible. Consequently one is obliged  to search for measurable quantities which are directly related to physical properties of interest.
It has been known for some time  \cite{surko69,donnelly80, walsh00} that the $1s^2 \rightarrow 1s2p(^1P)$ excitation of helium atoms confined and compressed is shifted to energies higher than in the free atom and that this energy shift depends on the helium density, $n$.
Both absorption and Raman spectroscopy studies of bulk helium  have provided energy shift values at known densities  and pressures\cite{surko69,arms01,arms05}. 
In [\!\!\citenum{pyper01}]  non-empirical electronic structure calculations for bulk compressed helium were presented
which  correctly reproduced the few experimental results available in 1999.
In the 2000s, Arms \etal\cite{arms01,arms05} presented a much more extensive set of energy shift measurements as a function of independently determined densities. These results were in agreement with the predictions of Pyper {\it et al.}\cite{pyper01}.
Further, using the same theoretical approach as in [\!\citenum{ pyper01}]
good agreement was found\cite{pyper21} for the full range of the bulk data of Arms {\it et al.}.

In recent decades, transmission electron microscopy has undergone significant advancements in both imaging techniques and analytical methods. These developments have facilitated the study of small objects with spatial resolutions at the nanometer and sub-nanometer scales. The use of scanning transmission electron microscopy (STEM) in conjunction with electron energy loss spectroscopy (EELS) now facilitates study in fine detail the properties of helium aggregates.
In these studies, the energy shift and the distance $h\sr{exp}$ traversed through just the helium by the incident electron beam are derived from experiment along with the values of the scattered beam intensities.  These latter are the intensity $I\sr{ZLP}$ of the beam emerging from the pore and encapsulating material without energy loss plus the corresponding  intensity $I\sr{He}$ of the beam which has lost energy through helium atom excitation.
  The helium number density can then, in principle, be determined from the relation\cite{walsh00, egerton11}:
\begin{eqnarray}\label{one}
n = \frac{ I\sr{He} }{ \sigma^{2p}(n, E_0) I\sr{ZLP} h_{exp}}.
 \end{eqnarray} 
Here, $\sigma^{2p}(n, E_0)$ is the theoretical  scattering cross-section which can be derived from electronic structure computations.  This  depends on both the helium density and the energy $E_0$ of the incident electron beam. The experimentally determined 
quantities are $I\sr{ZLP}$, the integrated intensity of the elastic peak,  and $I\sr{He}$, the intensity of the peak corresponding to the $1s^2\rightarrow1s2p(^1P)$ transition in helium, conventionally known as the K-edge.

Unfortunately, in the vast majority of attempts \cite{walsh00,taverna08,frechard09,david11,alix15,ono19,evin21,taylor22} to deduce the density from (\ref{one})
it was assumed that the scattering cross-section of an atom in the bubbles was essentially the same as that of a free helium atom, $\sigma^{2p}(0, E_0)$.
However, it has been shown\cite{pyper22} that the cross-section $\sigma^{2p}(n, E_0)$ is very strongly density dependent. It follows that  the density  is not explicitly determined by (\ref{one}). It is rather only given implicitly since $\sigma^{2p}(n, E_0)$ is itself a function of density.

It has recently been shown\cite{pyper22,pyper23}, that to make further progress, the experimental results must be analysed by simultaneously considering both the directly measured energy shifts $\Delta E\sr{obs}$ and the combination, to be denoted $\Sigma\sr{exp}(E_0)$, which is dependent on the purely experimental quantities $I\sr{ZLP}, I\sr{He}$ and $h\sr{exp}$ according to
\begin{eqnarray}\label{eqn:Sigma_exp}
\Sigma\sr{exp}(E_0)=\frac{ I\sr{He} }{  I\sr{ZLP} h\sr{exp}}.
\end{eqnarray}
Experimental data\cite{david11} for helium bubbles encapsulated in solid silicon subsequent to ion implantation have been analysed\cite{pyper23} using this approach to show that the behaviour of the helium in the bubbles was essentially the same as that of bulk condensed phase helium. This result enabled the helium densities to be deduced, thus enabling the pressures to be derived using a standard equation of state.
A further complication arises in the case of pores. Since they are manifestly non-spherical (see Figure \ref{fgr:Fig2}), the determination of ${h}\sr{exp}$ is more demanding.

This paper has three objectives. The first is to determine the  characteristics  of helium confined in pores in amorphous silicon. The second is to elucidate the relation between helium confined and compressed in the bulk, in bubbles and in pores.The third is to introduce a new way of using EELS measurements  to determine pore depths.
\section{Experiment}\label{exp}
\subsection{Motivation}\label{exp_motiv}
 In a prior experimental investigation  \cite{schierholz15}, EELS was employed in conjunction with STEM to study the state of helium condensed within small but non-spherical pores formed in amorphous silicon thin films.
Unfortunately, for the helium 1s$^2$ to 1s2p$(^1P)$ excitation cross-section, it was assumed\cite{schierholz15}, both that, the density independent free atom value could be used in (\ref{one}) and that this could be computed using the code Sigmak3.  However, this code yields not the excitation cross-section, but the hydrogenic ionization cross-section.  
 These considerations show that
 a thorough re-examination  of these results and data in  [\!\!\citenum{schierholz15}] is warranted.
Experimental data from the same study for pores not previously considered in [ \!\!\citenum{schierholz15}] are also analysed here.

In [\!\!\citenum{schierholz15}], helium filled pores of non-spherical shape, embedded and distributed over the whole thickness of amorphous silicon thin films, were produced using magnetron sputtering deposition in a helium atmosphere.  Such a bottom-up approach contrasts with conventional ion implantation top-down approaches reported in [\!\!\citenum{david11,david14,alix15}] that result from bombarding bulk crystalline silicon samples with high energy helium ions to create spherical helium bubbles distributed over a controlled depth below the surface.
The present reanalysis of the experiments\cite{schierholz15} enables the behaviour of the helium in these two environments to be compared, both with each other, and with that of the condensed phases of bulk helium.
\subsection{The EELS Spectra}
\begin{figure}[h!]
  \includegraphics[scale=0.66]{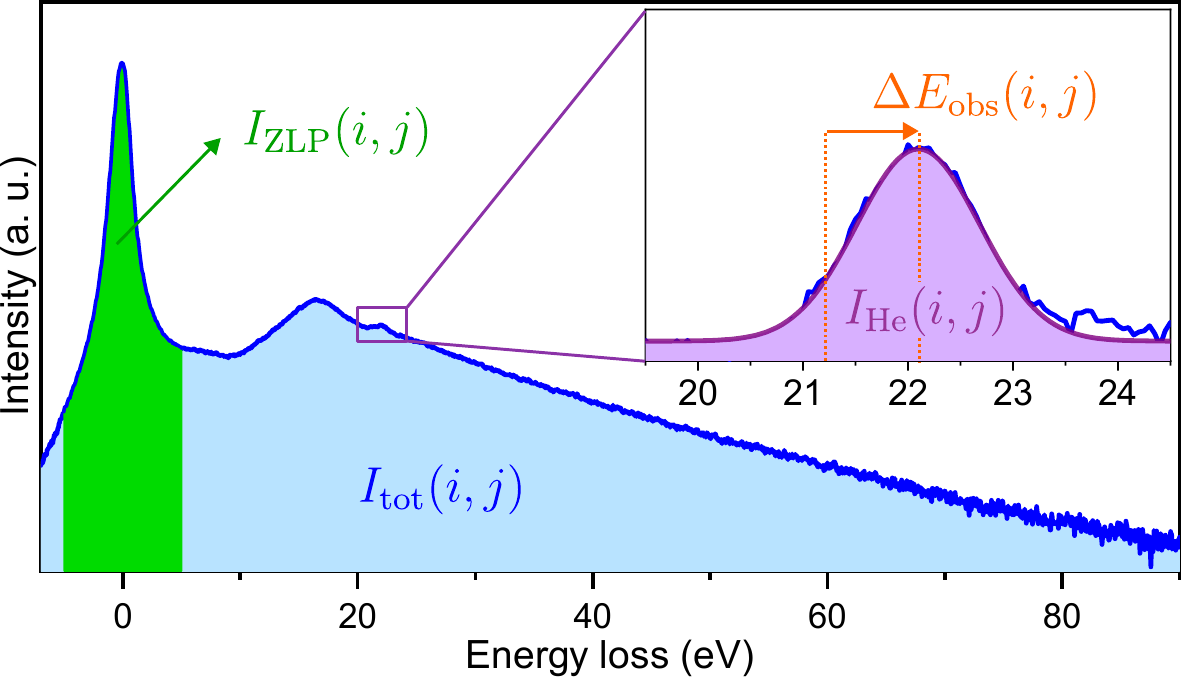}
\caption{\it{A typical recorded spectrum for a single pixel $(i,j)$.  $I\sr{ZLP}(i,j)$ and $I\sr{He}(i,j)$ are the intensities of the zero energy loss and inelastically scattered peaks respectively. The density-dependent energy shift, $\Delta E\sr{obs}(i,j)$ was obtained from the fitting of the He-K edge. $I\sr{tot}(i,j)$ is the total intensity of the whole EELS spectrum in arbitrary units.}}
 \label{fgr:Fig1}
\end{figure}

Helium  pores were generated during a thin film deposition process of a silicon sample via magnetron sputtering using helium as the process gas following the procedure in [\!\!\citenum{schierholz15}] and [\!\!\citenum{godinho13}].
The sample was then imaged by STEM using a high-angle annular dark-field (HAADF) detector to determine the location of isolated pores within the sample.
In this work, four isolated pores scanned according to the following procedure, were analysed in detail.
First, a HAADF survey image was taken of a pore and its surroundings.
The selected pore was then scanned in detail with a small electron probe ($\sim$ 2 \angstrom), using spatial steps of $\sim$ 1 nm, while simultaneously recording EELS signals up to 90\,eV. The spatially-resolved dataset consists of a grid of pixels (of approximately 1 nm by 1 nm in size) in the $x,y$ plane perpendicular to the direction $z$ of the incoming beam. The $z$ coordinate is defined to be zero at the top of the sample and to increase with distance into the material to become equal to the sample depth ($t$). The beam traversing any pixel exits the sample at the positions for which $z=t$, these being the only points at which the beam intensities can be measured. The quantities observed for each pixel are therefore labelled $(i,j)$.
  For each pixel, a separate EELS spectrum was stored and the EELS data was then processed according to the protocol detailed in [\!\!\citenum{schierholz15}], which
encompasses spectra alignment, noise reduction, and deconvolution of multiple scattering effects.  Figure \ref{fgr:Fig1} shows the EELS spectrum for one such pixel.  
\FloatBarrier
The following quantities were derived from the analysis of the EELS spectrum:
\begin{itemize}
\item the total collected intensity, $I\sr{tot}(i,j)$, being the integral of the collected signal from about $-10$eV to $90$eV;
\item the integrated intensity of the zero-loss peak, $I\sr{ZLP}(i,j)$, being the integral of the collected signal from $-5$eV to $5$eV;
\item the integrated intensity of the excited helium peak, $I\sr{He}(i,j)$, determined as the integral of a gaussian (whose centre lies in the 20--24\,eV) fitted to the residual remaining after subtracting the background silicon signal; and
\item the directly observed helium peak energy shift, $\Delta E\sr{obs}(i,j)$, being the difference between the centre of the fitted gaussian shown above and the reference free helium excitation value of 21.218 eV.
\end{itemize}

\begin{figure}
  \includegraphics[scale=0.77]{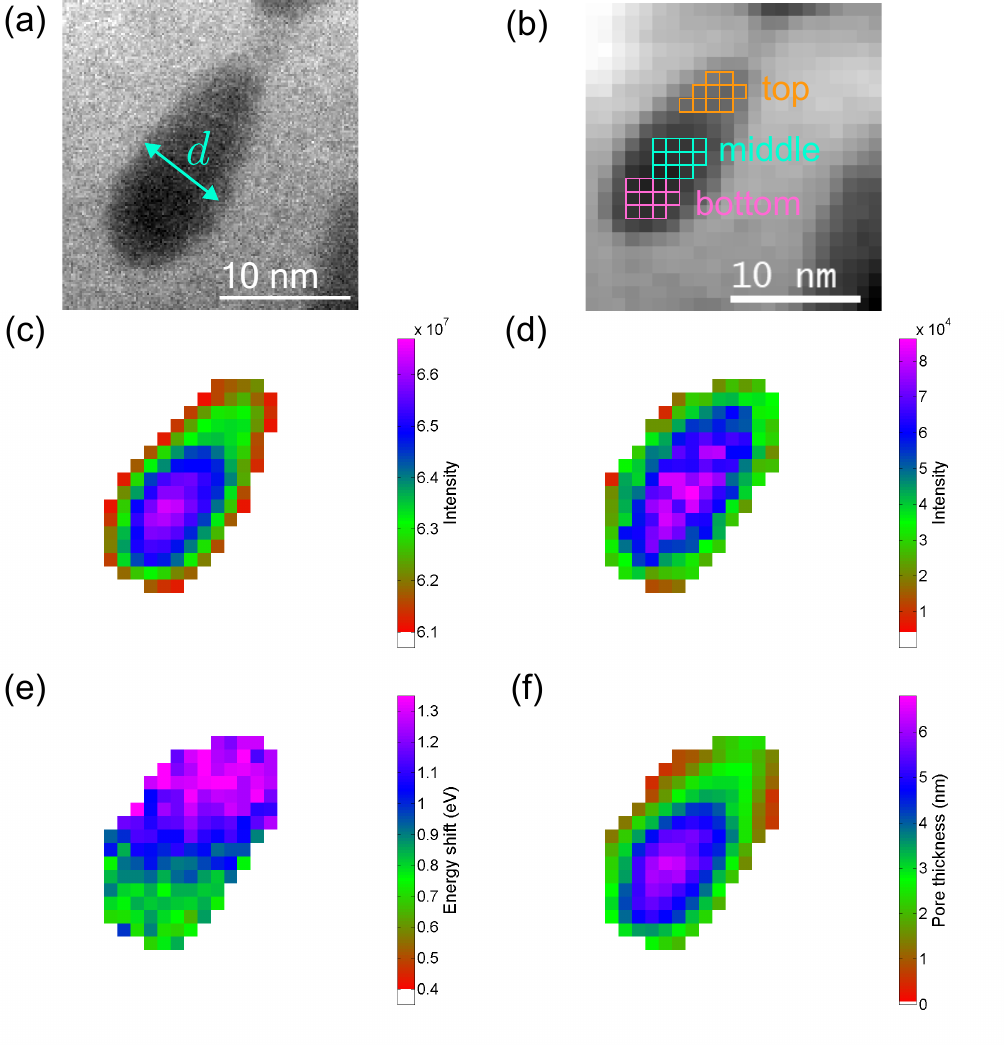}
  \caption{\it{Spatially-resolved maps extracted of pore 1 from the STEM-EELS analysis (see text for more details): (a) and (b) show the HAADF signal recorded before and during the STEM-EELS acquisition, respectively.  For reference, the width of the pore measured at its middle (as in (a)) is 7.0\,nm. Small coloured squares indicate the pixels that have been selected for the analysis in each pore region (``top'', ``middle'' and ``bottom''). 2D maps of the most relevant extracted parameters: (c) $I\sr{ZLP}(i,j)$, (d) $I\sr{He}(i,j)$, (e) $\Delta E\sr{obs}(i,j)$ and (f) $h(i,j)$. For each pixel, the value of $h(i,j)$ was deduced from the corresponding values of the electron beam intensities using the theory presented in Section \ref{sec:det_h}}}
  \label{fgr:Fig2}
  \end{figure}

The information for each pixel that has been extracted from the STEM-EELS analysis of Pore 1 is displayed in Figure \ref{fgr:Fig2}. The HAADF survey image (as shown in Fig. \ref{fgr:Fig2}a) shows the pore (in dark contrast) surrounded by the silicon matrix (in bright contrast). Using this image, it is possible to ascertain the projected diameter $d$ of the pore perpendicular to its major axis by utilizing an intensity profile.
 Fig. \ref{fgr:Fig2}b shows the corresponding HAADF which was recorded  simultaneously with the EELS scan. 
The STEM scanning proceeded in a left-to-right and top-to-bottom manner across the displayed data. For Pore 1, the corresponding 2D maps for $I\sr{ZLP}(i,j)$, $I\sr{He}(i,j)$, and $\Delta E\sr{obs}(i,j)$ are shown in Fig. \ref{fgr:Fig2}c, Fig. \ref{fgr:Fig2}d, and Fig. \ref{fgr:Fig2}e respectively.
From these experimental quantities, it is possible to determine an estimate of the depth, $h(i,j)$ of the pore at pixel (i,j), using an approach described in section \ref{sec:det_h}.  A two-dimensional map of these depths is plotted in Fig. \ref{fgr:Fig2}f.
It is to be noted that the extraction of $\Delta E\sr{obs}(i,j)$ from the EELS spectra in regions outside the pore or at its periphery may yield arbitrary values lacking physical significance due to the inability to accurately fit the He peak owing to the absence of a strong He signal.  Consequently, to enhance clarity, the background area surrounding the pore has been eliminated in each map.

  \FloatBarrier
\subsection{Variations between portions of pores}
The spatially resolved pixels maps in figure \ref{fgr:Fig2} for pore 1 show that all the three measured quantities, $I\sr{ZLP}(i,j)$, $I\sr{He}(i,j)$, and $\Delta E\sr{obs}(i,j)$ vary significantly within the pore.  The data for the three other pores all show similar variations. However, for both pores 1 and 2, one can identify three different spatial regions each composed of sets of contiguous  pixels in which the values of $I\sr{ZLP}(i,j)$, $I\sr{He}(i,j)$, $\Delta E\sr{obs}(i,j)$ do not vary significantly. For each of pores 3 and 4, there are two such regions within which these three quantities are essentially constant.

%
These \quotes{pore portions} are designated as \quotes{top}, \quotes{middle} or \quotes{bottom}, and for Pore 1 are shown in Fig. \ref{fgr:Fig2}b.
The top portion gave the largest energy shift in each case. This can be  understood as a degassing effect due to the interaction of the electron beam  with the target. Such degassing has been observed and studied in great detail for encapsulated helium bubbles in silicon\cite{david14}.  As the density of helium in the pore is reduced the energy shift will be diminished as predicted in [\!\!\citenum{pyper22}].

For each pore portion, the measured values of $I\sr{ZLP}(i,j), I\sr{He}(i,j)$, and  $\Delta E\sr{obs}(i,j)$ were averaged over the pixels, to produce values designated by the absence of the (i,j) labels.  The depth of each pore portion, designated $h\sr{exp}$ was derived as the average of all the $h(i,j)$.
It is these averaged results that appear in equation (\ref{eqn:Sigma_exp}) which are used for this analysis.
\section{Theoretical Analysis of Experimental Data}\label{sec:theory}
\subsection{General Theory of Scattering Through a Material}



An energetic electron beam passing through a material will be attenuated due to inelastic collisions with the atoms in the material. If the material has a thickness of $\tau$ and the incoming beam has 
an intensity of $I\sr{in}$ then the exiting beam will have an intensity
\begin{eqnarray}\label{colm1}
I\sr{out}= I\sr{in}\exp\left(-\frac{\tau}{\lambda\sr{mat}}\right),
\end{eqnarray}
where $\lambda\sr{mat}$ is the mean free path for scattering in the material.  For a material composed of weakly interacting, but clearly identifiable atoms, $\lambda\sr{mat}$ can be related to their number density $n\sr{mat}$ and the inelastic cross-section $\sigma(n\sr{mat},E_0)$ of one such atom by
\begin{eqnarray}\label{colm2}
n\sr{mat}\sigma(n\sr{mat},E_0)=\frac{1}{\lambda\sr{mat}}.
\end{eqnarray}
A more detailed discussion\cite{egerton11} shows that results of the type (\ref{colm1}) are only valid if the scattering length is much greater than the thickness of the sample. It is only under these conditions that the scattering behaviour is Poisson-like with multiple scattering events having very small probabilities.
\subsection{Determination of $\Sigma\sr{exp}$}
\begin{figure}[t!]
 \centering
 \textbf{Diagram for relating theoretically derived properties of encapsulated helium to the experimentally measured intensities entering the expression $\Sigma\sr{exp}(E_0)$}\\

 \includegraphics[scale=0.75]{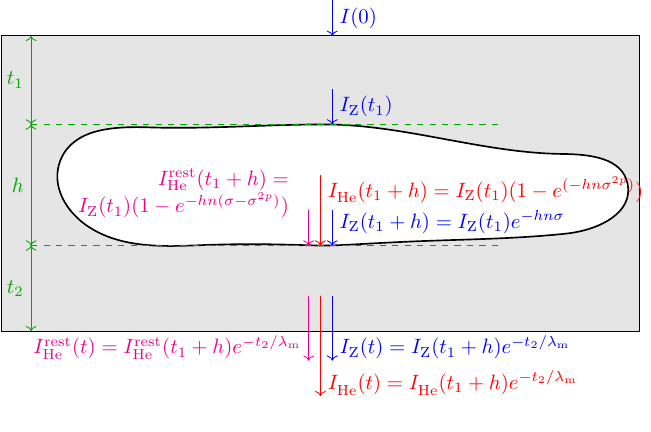}
 \caption{\it{The electron beam intensity through the specimen thickness varies due to scattering events in the matrix and in the He pore.
 At a given position in the specimen where there is a single pore present, the thickness of the matrix above the bubble, the depth of the bubble and the thickness of the matrix below the bubble are denoted $t_1$, $h$ and $t_2$ respectively;  this corresponds to a single pixel at position $(i, j)$, but these coordinates are omitted for clarity.
 As the beam progresses through the sample at this position, the unscattered intensity is labelled $I\sr{Z}(z)$, and the intensity {of the inelastically scattered electron beam resulting in helium excitation to the \twop\ state is} denoted $I\sr{He}(z)$. $I_{He}^{rest}(z)$ is the total  intensity  scattered into levels higher in energy than the $1s 2p({}^1P)$  state.
 The quantities given on right of the figure are the beam intensity ($I(z)$) at respective planes in the sample labelled by their $z-$depth within the sample, being (top to bottom) the top of the sample ($z=0$), the top of the pore ($z=t_1$), the bottom of the pore ($z=t_1+h$), and the bottom of the sample ($z=t_1+h+t_2$, labelled $(t)$ as these are experimentally measured).
  The mean free path for electron scattering by the matrix is denoted $\lambda\sr{m}$.
\label{colmfiga}}}
\end{figure}

In the current experiments there are
two different materials: an encapsulating silicon matrix with associated mean free path $\lambda\sr{m}$  and the helium-filled pore  with mean free path
$\lambda\sr{He}$. 
The electron beam is progressively scattered as it passes through the material.  To analyse this behaviour,  we retain the $z$ coordinate, indicating a depth through the material, and denote the electron beam intensity there as just $I(z)$ (see Figure \ref{colmfiga}).
The pore  is encapsulated in the film a distance $t_1$ from the top of the film, 
with the distance through the pore along the path traversed by the beam being denoted by $h$. Since the theory is the same for all the pixels belonging to each individual pore portion, the $(i,j)$ labels will be omitted from $h$, $\Delta E\sr{obs}$ and the intensities, with each of these quantities being designated by just the appropriate value of the $z$ coordinate.  
{It is important to distinguish the attenuation of the beam due to scattering (with cross-section $\sigma(n, E_0)$) into all possible excited states from the scattering (with cross-section $\sigma^{2p}(n,E_0)$) into just the $1s2p(^1P)$ state.
  The total cross-section for inelastic scattering events in the helium gas can be expressed as the sum of the cross-sections for each transition\cite{walsh00}.  
}

The beam enters the film with an intensity $I(0)$ 
{to impinge on the bubble with its intensity reduced by the factor $\exp(-t_1/\lambda\sr{m})$ to the value $I\sr{Z}(t_1)$, with the value $I(0)$ not being relevant to the analysis in this subsection.}
 As shown in figure \ref{colmfiga}, inelastic scattering by the helium in the pore portion reduces the intensity of the elastically scattered beam from $I\sr{Z}(t_1)$ on entering the pore to $I\sr{Z}(t_1+h)$ at the exit point.
{Combining relations (\ref{colm1}) and (\ref{colm2}) and temporarily dropping the dependence of the cross-section on $n$ and $E_0$ shows that these intensities are related by 
\begin{eqnarray}\label{colm4}
I\sr{Z}(t_1+h)&=& I\sr{Z}(t_1)\exp\left(-hn\sigma\right).
\end{eqnarray}
Here, the inelastic scattering probabilities are sufficiently small compared to the elastic scattering probability that the scattering events into different states can be treated as essentially independent.
On leaving the pore at $z$ coordinate $t_1+h$, the intensity of the beam inelastically scattered {into the \twop\ state} will be the difference between the intensities of the elastically scattered beam at the entrance and exit positions in the pore, so that
\begin{eqnarray}
I\sr{He}(t_1+h)&=&I\sr{Z}(t_1)\left[1-\exp(-hn\sigma^{2p})\right]\label{colm7}
\end{eqnarray}
Dividing (\ref{colm7}) by (\ref{colm4}),
\begin{eqnarray}\label{colm9}
\frac{I\sr{He}(t_1+h)}{I\sr{Z}(t_1+h)}= \frac{I\sr{Z}(t_1)\left[1-\exp(-hn\sigma^{2p})\right]}{I\sr{Z}(t_1)\exp(-hn\sigma)}= \frac{1-\exp(-hn\sigma^{2p})}{\exp(-hn\sigma)}=\exp\left(hn\sigma\right)-\exp\left[hn(\sigma-\sigma^{2p})\right].
\end{eqnarray}
For a typical helium bubble/pore density of 0.05 atoms \colmang$^{-3}$ for which the scattering cross-section is around $4.2\times 10^{-4}\,$\colmang$^2$ for the 300\,keV incident energy used in the pore experiments, the product $n\sigma$ has value $2.17\times10^{-5}\colmang^{-1}$.
The independent determinations of pore depths given in Table \ref{tbl:h_exp} give $h<100\,\colmang$.
This shows that $hn\sigma$ and $hn(\sigma-\sigma^{2p})$ are sufficiently small that each exponential may be expanded  to first order to yield
\begin{eqnarray}\label{colm9}
\frac{I\sr{He}(t_1+h)}{I\sr{Z}(t_1+h)}= hn\sigma^{2p}(n,E_0),
\end{eqnarray}
where we have restored the dependence of $\sigma^{2p}$ on $n$ and $E_0$.
{This result extends the derivation of equation (4) of Walsh \etal\cite{walsh00} by explicltly considering more than one possible helium transition, and allowing for a non-spherical pore.}

 On travelling from the bottom of the pore at $z = t_1+h$ to exit at $z = t = (t_1+h) + t_2$, the intensities, $I\sr{Z}(t_1+h)$ and $I\sr{He}(t_1+h)$, of the beams both elastically and inelastically scattered by the helium in the pore are further attenuated by the same factor of $\exp(-t_2/\lambda\sr{m})$ as shown in Figure \ref{colmfiga}.  
This observation, after division through by $h$, allows (\ref{colm9}) to be written as 
\begin{eqnarray}\label{colm10}
\frac{I\sr{He}(t_1+h)}{hI\sr{Z}(t_1+h)}= \frac{I\sr{He}(t)}{hI\sr{Z}(t)}= n\sigma^{2p}(n, E_0).
\end{eqnarray}
Since each pore portion is defined as that collection of contiguous pixels for which the observed values $I\sr{He}(t), I\sr{Z}(t)$ and $h$ do not vary significantly, an equation similar to (\ref{colm10}) can be validly constructed from the averages $I\sr{He}, I\sr{ZLP}$ and $h\sr{exp}$.  This yields

\begin{eqnarray}\label{colm11}
\frac{I\sr{He}}{h\sr{exp}I\sr{ZLP}}= n\sigma^{2p}(n, E_0),
\end{eqnarray}
where $n$ is an average helium density within the pore portion.
The quantity $I\sr{He}/(h\sr{exp}I\sr{ZLP})$ is that defined as  the purely experimental quantity $\Sigma\sr{exp}(E_0)$ in equation (\ref{eqn:Sigma_exp}).
The result (\ref{colm11}) is independent of both $t_1$ and $t_2$. Thus the ratio is independent both of   the film thickness and  the position of the pore in the film.
Consequently the same analysis applies to all the pores studied here.

The equality between the experimental $\Sigma\sr{exp}(E_0)$ with the product of the purely theoretical quantities of the right hand side of equation (\ref{colm11}) indicates that $n\sigma^{2p}(n,E_0)$ is the exact theoretical analogue, to be denoted $\Sigma\sr{theory}(n,E_0)$, of $\Sigma\sr{exp}(E_0)$. Hence $\Sigma\sr{theory}(n,E_0)$ can be defined as
\begin{eqnarray}
\Sigma\sr{theory}(n, E_0)&=&n\sigma^{2p}(n, E_0).\label{eqn:Sigma_theory}
\end{eqnarray}
This theoretical quantity should equal\cite{egerton11,pyper22} the corresponding experimental quantity $\Sigma\sr{exp}(E_0)$ provided both that the experiments are essentially free of errors and that the $n\sigma^{2p}(n,E_0)$ have been calculated sufficiently accurately.

}

\subsection{Determination of pore depths from EELS measurements}\label{sec:det_h}
\FloatBarrier
\begin{figure}[t]
 \centering
 \textbf{Analysis of the two measurements needed to determine one pore depth}\\
 \includegraphics[scale=0.75]{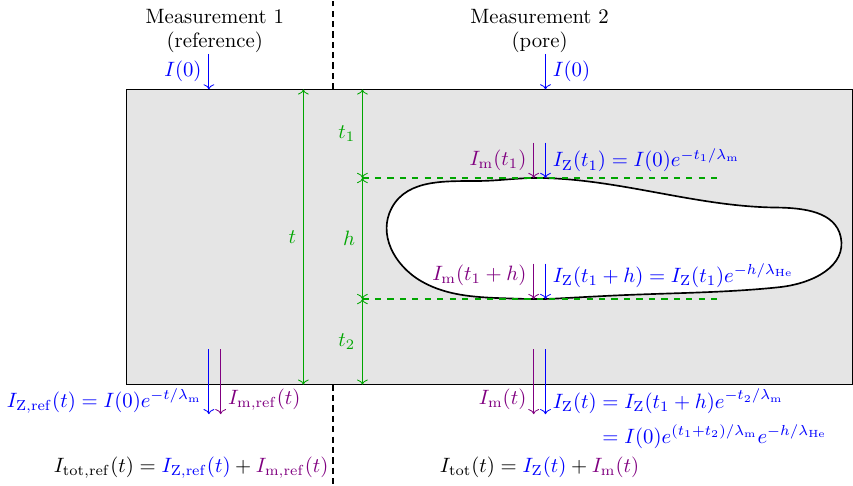}
 \caption{\it{
Two measurements are carried out to determine the thickness of the pore: Measurement 1 on a sample of matrix close to the pore (with quantities labelled with subscript `$,\mathrm{ref}$'); and Measurement 2 passing through the pore itself.
See caption to Figure \ref{colmfiga}}.
$I\sr{m}(z)$ is the intensity at the position, defined by $z$ of the electron beam inelastically scattered by the matrix.
\label{colmfigb}}
\end{figure}
In
previous studies of He bubbles\cite{walsh00,taverna08,frechard09,david11,alix15,ono19,evin21} , it was generally assumed that they are perfectly spherical so  that the depth $h$ of the bubble through the centre can just be taken to be $d$ the diameter of the circular projection of the bubble in the HAADF image\cite{walsh00,frechard09}.
In our case the HAADF image of the pore is distinctly non circular (Fig. \ref{fgr:Fig2}a), rather it has a \quotes{teardrop} or \quotes{tadpole} shape, with a  globular head at the bottom (which is actually closer  to the film surface) tapering to a narrow tail at the top (closer to the substrate).
For these novel morphologies a new approach is needed, as suggested by earlier authors, although specific details of the approach were not provided\cite{taverna08,david11}.

Given that the pores have irregular morphologies (tadpole /tear-like shapes), the assessment of their dimension $h$ along the electron trajectory poses a significant challenge and is likely to be the source of the highest uncertainties. In this case, the use of an approach based on the measurement of the projected width $d$ derived HAADF images, which constitutes a straightforward way for gauging the thickness of spherical nanobjects, is clearly unsuitable. 
The method employed here is  based on the log-ratio method of Egerton\cite{egerton11}. 
 Below we describe the method for a uniform material, and then the modifications needed to apply to the non-uniform material containing a He pore.

Figure \ref{colmfigb} depicts the path of the electron beam incident on pixel in a general section of the specimen which contains a He pore.  Two sets of measurements are taken.  The first, Measurement 1, is for a reference beam passing close to the pore but entirely through the uniform encapsulating material. The second measurements, shown in the right of Figure \ref{colmfigb} refer to a beam travelling through both the pore itself as well as parts of the encapsulating matrix.
Since the theory is the same for all the pixels, the $(i,j)$ labels will also be omitted.

In figure \ref{colmfigb}, the electron beam is progressively scattered as it passes through the material.  To analyse this behaviour, the $z$ coordinate is retained, indicating a depth through the material, with the electron beam intensity there denoted as just $I(z)$.
As the beam progresses and is scattered by the material, this intensity divides into: the unscattered intensity, $I\sr{Z}(z)$, the amount scattered by the matrix, denoted with subscript m as $I\sr{m}(z)$, plus that scattered by the helium, $I\sr{He}(z)$.

At the reference position show in Figure \ref{colmfigb} , the intensity $I(0)$ of the incident unscattered beam is diminished during its passage through the specimen due to inelastic scattering events caused by the silicon matrix, these being governed by Poisson statistics.
The measured intensity $I\sr{Z,ref}(t)$ of the unscattered beam exiting the sample is thus given by
\begin{equation}
I\sr{Z,ref}(t)=I(0)e^{-t/\lambda\sr{m}},\label{eqn:IZref}
\end{equation}
where $\lambda\sr{m}$ is the electron mean free path in the matrix.
However, in the rearrangement of (\ref{eqn:IZref}) to yield
\begin{equation}
t=\lambda\sr{m}\ln\left(\frac{I(0)}{I\sr{Z,ref}(t)}\right)\label{eqn:t_exact},
\end{equation} the quantity $I(0)$
 cannot be directly measured, but can be approximated as the total intensity of the whole EELS spectrum $I\sr{tot,ref}(t)$.
This enables the reference thickness to be calculated from experimentally measured quantities:
\begin{equation}
  t =\lambda\sr{m}\ln\left(\frac{I\sr{tot, ref}(t)}{I\sr{Z, ref}(t)}\right). \label{eqn:log_ratio2}
\end{equation}

In the present study, the examined material exhibits non-uniform characteristics, wherein helium pores are embedded within a silicon matrix. As illustrated in Fig. \ref{colmfigb}, the incident beam intensity diminishes during its passage through the specimen due to inelastic scattering events occurring within both the matrix and the helium condensates.
Similarly to the result (\ref{eqn:IZref}), the intensity of the unscattered beam at the exit side of the specimen, can be expressed as \cite{walsh00}:
\begin{equation}
  I\sr{Z}(t) = I(0) e^{-t\sr{m}/\lambda\sr{m}}  e^{-h/\lambda\sr{He}}, \label{eqn:Iz}
\end{equation}

\noindent where $t_{m}$ is the total thickness of the matrix above and below the pore (corresponding to $t\sr{m}=t_1+t_2$ in Fig. \ref{colmfigb} ) and $\lambda\sr{He}$ is the mean free path associated with inelastic scattering events within the helium volume.
Following the analysis of the previous section, $\lambda\sr{He}$ has a value of approximately 46000\colmang. This shows that the contribution that $e^{-h/\lambda\sr{He}}$ makes to the difference between $I\sr{Z}(t)$ and $I(0)$ in (\ref{eqn:Iz}) is negligible compared with that generated by the term $e^{-t\sr{m}/\lambda\sr{m}}$ arising from the inelastic scattering by the silicon matrix. Consequently $t_m$ can be calculated from

\begin{equation}
t\sr{m}=\lambda\sr{m} \ln\left(\frac{I\sr{tot}(t)}{I\sr{Z}(t)}\right).\label{eqn:t_m}
\end{equation}
The pore thicknesses, $h$, of the material at each pixel (see Fig. \ref{fgr:Fig2}f) are calculated as the complement of the matrix thickness as $h=t-t\sr{m}$ by combining (\ref{eqn:log_ratio2}) and (\ref{eqn:t_m}) as
\begin{equation}
h=\lambda\sr{m} \ln\left(\frac{I\sr{tot, ref}(t)I\sr{Z}(t)}{I\sr{Z,ref}(t)I\sr{tot}(t)}\right).\label{eqn:h_calc}
\end{equation}
For each of the ten pore portions, the final depth, denoted $h\sr{exp}$ was calculated by averaging the predictions of(\ref{eqn:h_calc}) over all the constituent pixels

 Evaluation of the distances $t$ and $t\sr{m}$ requires a value for the silicon scattering length $\lambda\sr{m}$. For variety of materials but not including silicon, values of these lengths have been derived\cite{malis88} through equation (\ref{eqn:log_ratio2}) after having determined the thicknesses ($t$) of each sample from entirely different independent experimental measurements.
 The results of these measurements leave $\lambda\sr{m}$ as the only remaining unknown, this then being calculated via (\ref{eqn:log_ratio2}).
 The values of the $\lambda\sr{m}$ depend ultimately on the nuclear charge and impact energy $E_0$.
 The dependence of the set of $\lambda\sr{m}$ values on these variables has been made explicit by the relation (7) of Ref. \!\!\citenum{malis88}.
 It has been estimated\cite{malis88} that errors of up to 20\% can arise when using this relation to calculate $\lambda\sr{m}$ values for materials not in the original data set.
 For silicon, direct use of this relation yields a value of 1463$\colmang$  for the $E_0=300\,\keV$ used in the present study.
  The slightly different value of $\lambda\sr{m}=1447 \colmang$ used in (\ref{eqn:t_exact}) above was derived from the $1463 \colmang$ prediction by introducing the convergence correction described by Egerton (Ref. \!\!\citenum{egerton11} p. 276).
\section{Systematic Experimental Error in $\Delta E\sr{obs}$}
The quantities derived from or directly measured in the STEM/EELS experiments are $\Sigma\sr{exp}(E_0)$ and values for the $1s^2 \rightarrow1s2p(^1P)$ excitation energy.  The energy shift can,  at least in principle, be derived by subtracting the free atom excitation energy of 21.218eV (see Figure \ref{fgr:Fig1}). The STEM data can then be analysed from graphs of energy shift plotted against $\Sigma\sr{exp}(E_0)$. 
 However,  the observed K-edge is sensitive  to  the resolution  and  other parameters specific to a given  microscope \cite{egerton11,alix15, alix18}.
 In their  re-examination of their earlier  data\cite{david11,david14,alix15} the Poitiers group identified a systematic error in the measured $1s^2\rightarrow 1s2p(^1P)$ excitation energies  and the shifts derived from them were corrected by a factor $\delta E^s$ determined by requiring that the least dense bubble should have a zero energy shift.  In  [\!\! \citenum{pyper23}] it was shown that this leads to an overestimation of the systematic error and  a better estimate  was provided.
A preliminary examination of the current STEM measurements indicate that such a systematic error might have arisen.
Following exactly the same procedure as in [\!\! \citenum{pyper23}], it is noted that
in the limit of zero density (the free atom case) one must have
\begin{eqnarray}\label{limit}
\lim_{n\rightarrow 0}\Sigma\sr{theory}(n, E_0)=\lim_{n\rightarrow 0}n\sigma^{2p}(n, E_0)=0.
\end{eqnarray}

At a higher density  of 0.022 \dena, the helium is liquid and  experiences its ambient vapour pressure\cite{surko69}.  For this system, the experimentally determined pair distribution function showed that the environment of each atom, in its first two coordinating shells, was similar to a body centred cubic lattice. As depicted in figure 6 of
 [\!\!\citenum{lucas83}], the first coordination shell of eight atoms is clearly shown, but the second coordination shell was much less well defined.
Consequently, in  [\!\!\citenum{pyper23}], the liquid at the even lower density of 0.0083 \dena was modelled by considering its interactions with only the two nearest shells of a bcc coordinating lattice.
This is justified by the excellent agreement between previous theoretical calculations and the experimental value for the low density liquid helium under its own vapour pressure\cite{pyper21,pyper23} 
For this arrangement of atoms, for a density of 0.0083 \dena, non-empirical electronic structure calculations\cite{pyper23} gave an excitation energy shift of 0.13eV.
At this low density, the cross-section is that of the free atom, with a value\cite{ral08} of $0.2665\times 10^{-3}\colmang{}^2$ 
 resulting in a $\Sigma\sr{theory}(0.0083\,\mathrm{atoms}\,\colmang^{-1}, 300\,\mathrm{keV})\approx0.22\times 10^{-5}$ atoms  $\colmang{}^{-1}$.
This theoretical low-$\Sigma\sr{theory}(n, E_0)$, low-energy-shift point will be used as a reference in assessing any systematic errors in $\Delta E\sr{obs}$.

\begin{figure}[t!]
\includegraphics[scale=0.5]{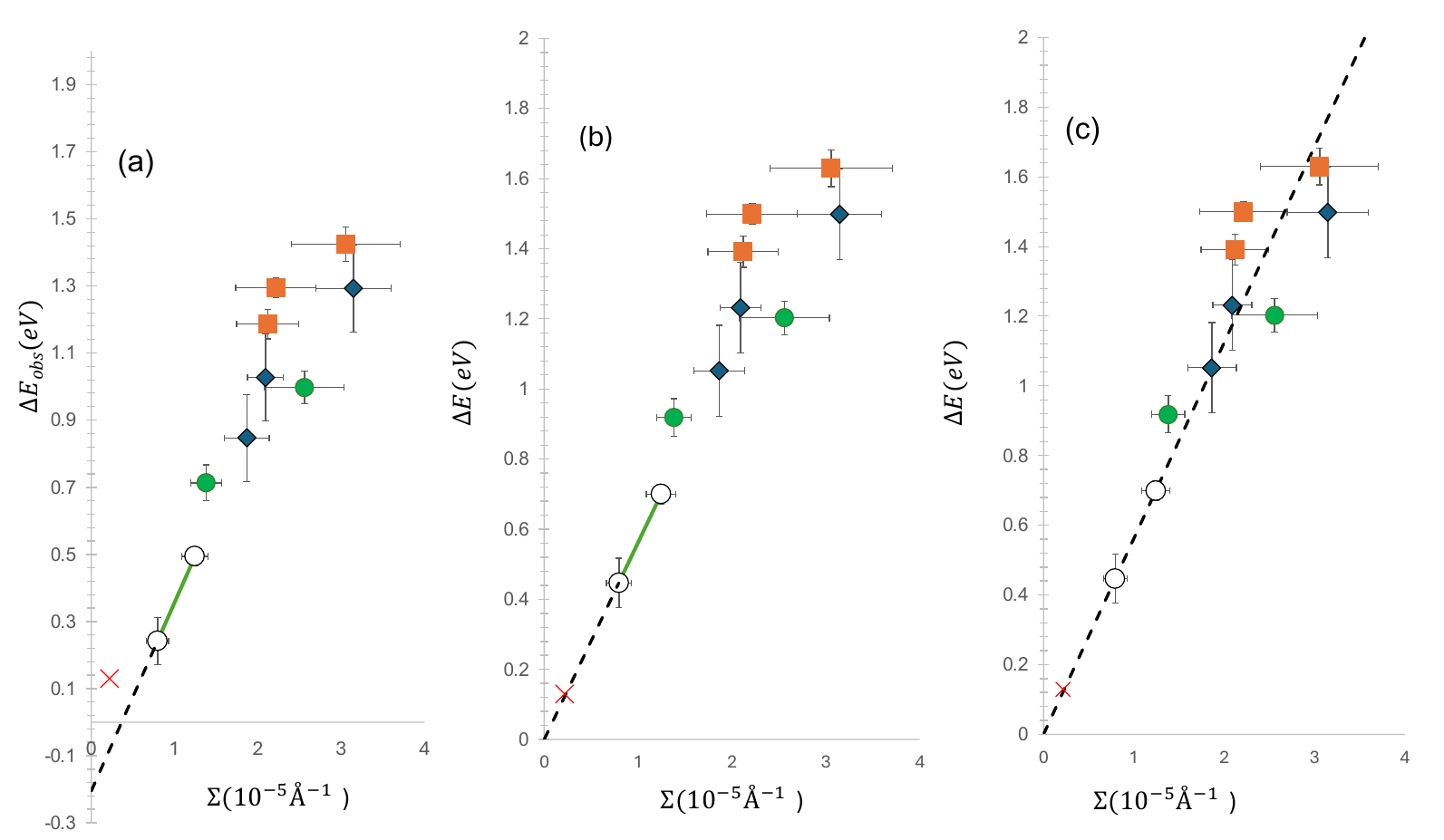}
\caption{ \it{Energy shifts plotted against  $\Sigma\sr{exp}(300\,\keV)$ for the different pores: blue-filled diamond-Pore1;
orange filled square-Pore 2; Open disk-Pore 3; green filled disk-Pore 4; the red cross is the reference point.  
The green line in both panels (a) and (b) connects the two points (both from Pore 3) with the smallest energy shifts (lowest densities) with each line being extrapolated to $\Sigma\sr{exp}(300\,\keV)=0$.
Panel (a)  presents the results using the directly measured shifts $\Delta E\sr{obs}$ shown in figure \ref{fgr:Fig1}.
In panels (b) and (c) the energy shifts $\Delta E$ are those ($=\Delta E\sr{obs}+0.205\,\text{eV}$) corrected to remove the systematic error.
In panel (c) the line through the Pore 3 points has been extended from the origin to cover the whole range of available data.
\label{systematic}}}
\end{figure}
\FloatBarrier

Figure \ref{systematic}a presents a graph of the directly measured energy shifts ($\Delta E\sr{obs}$) plotted against the $\Sigma\sr{exp}(E_0)$.
The two points at lowest density (Pore 3) are joined by a straight line,
which was extrapolated
to $\Sigma\sr{exp}(300\,\keV)=0$. This line  intersects the  $\Delta E\sr{obs}$ axis at a point $-0.205$ eV and it lies below the reference point.
This analysis suggests that the systematic error correction should be taken to be $\delta E^s=0.205$. In the energy shift v. $\Sigma\sr{exp}(300\,\keV)$ graph shown in panel \ref{systematic}b, the directly observed shifts $\Delta E\sr{obs}$ have been corrected by adding $0.205$ eV to produce values to be denoted $\Delta E$. A \quotes{best fit} line to  the two low density points now passes through the origin and the reference point, enabling it to be concluded that the corrected values $\Delta E$ are the most reliable results for the energy shifts.  These will therefore be used throughout this paper.

If this linear fit is further extended (Figure \ref{systematic}c), it is  found to be consistent with current experiment over the full range of data.
The equation of this line  is
\begin{eqnarray}\label{linearNoArms}
\Sigma\sr{fit}(300\,\keV)&=& \mu_{300}\Delta E
\end{eqnarray}
for $0\le\Delta E\le 1.3$ eV.  When $\Sigma\sr{fit}(300\,\keV)$ is expressed in $\colmang^{-1}$ with $\Delta E$ in eV, then $\mu_{300}=1.7801\times 10^{-5} \colmang^{-1} \text{eV}^{-1}$. In Figure \ref{systematic}c the current experimental results for $\Sigma\sr{exp}(300\,\keV)$ are shown as a function of the $\Delta E$. The line through the lowest energy shift points (\ie Pore 3) is extended  down to the origin and up to $\Delta E=1.8$ eV.

\section{Cross sections}

Calculation of the scattering cross-sections entering $\Sigma\sr{theory}(n, E_0)$ requires the electronic wavefunctions for the ground and $1s2p(^1P)$ excited states of helium atoms in bulk condensed phases. The computation of these density dependent Hartree--Fock wavefunctions has been fully described elsewhere\cite{pyper01}.
For both  the bubble and pore experiments the impact energy is sufficiently high that  the Born approximation\cite{inokuti71,pyper22} can be used, where, in atomic units,

 \begin{eqnarray}\label{born1}
\sigma^{2p}(n, E_0)&=&\frac{4\pi}{E_0}\int_{q\sr{min}}^{q\sr{max}}\frac{|\epsilon_{2p}(q)|^2}{q^3}dq.
\end{eqnarray} 
$q\sr{min}$ and $q\sr{max}$ are the minimum and maximum values of the momentum transfer, $q$, as given in [\!\!\citenum{pyper22}] and  $E_0$ is the energy in the incident electron beam.  $\epsilon_{2p}(q) $ is the  atomic form  factor which depends on the wavefunctions of the initial and final states of the target atom. The cross-section is dependent on the impact energy, both explicitly, and implicitly, through the energy dependence of the integration limits $q\sr{min}$ and $q\sr{max}$ in (\ref{born1}).
For beam energies of both 200\,keV and 300\,keV Table I presents, for all the bulk helium densities measured by Arms \etal.\cite{arms01,arms05}, the scattering cross-sections calculated using (\ref{born1}). 

\begin{table}[h!]
\caption{Cross-sections, $\sigma^{2p}(n, E_0)$, in $10^{-3}a.u.^2$ for different bulk He densities using the fcc extended model\cite{pyper01,pyper22} with impact energies of $E_0=200\,\keV$ (taken from Ref. \!\!\citenum{pyper23}), and recalculated for $E_0=300\,\keV$ using equation (\ref{born1}). \label{ratio}}
\begin{tabular}{|c|c|c|c|}
\hline
$n$(atoms $\colmang^{-3}$)      &$\sigma^{2p}\sr{fcce}(n, 200 \mathrm{keV})$&$\sigma^{2p}\sr{fcce}(n, 300 \mathrm{keV})$&$\frac{\sigma^{2p}\sr{fcce}(n, 200 \mathrm{keV})}{\sigma^{2p}\sr{fcce}(n, 300 \mathrm{keV})}$\\
\hline

0.0400& 2.0736  &1.3879&1.4940\cr\hline
0.0419&	2.1134	&1.4147&1.4939\cr\hline
0.0446&	2.1194	&1.4189&1.4937\cr\hline
0.0450&	2.1239	&1.4219&1.4937\cr\hline
0.0460&	2.1656	&1.4499&1.4936\cr\hline
0.0492&	2.2194	&1.4861&1.4934\cr\hline
0.0500& 2.2334	&1.4955&1.4934\cr\hline
0.0518&	2.2504	&1.5070&1.4933\cr\hline
0.0536&	2.2651	&1.5170&1.4932\cr\hline
0.0562&	2.3190	&1.5532&1.4932\cr\hline
\end{tabular}
\end{table}


\section{Pores, Bubbles and Bulk Condensed Helium}\label{sec:pores_etc}

\subsection{The Bulk Condensed Phases}
For both the studies of bubbles and pores, the determination of the density is crucial. 
In bulk helium it has been shown \cite{pyper21,pyper22} that there is a one-to-one relationship between energy shift and the density.
It has been, further, established\cite{pyper22} that the behaviour of the encapsulated helium in the bubbles is consistent with that in the bulk phase. In this section the behaviour of the encapsulated helium in the pores is considered and compared with that in the bulk and bubbles.

For bulk helium under accurately experimentally determined pressures sufficiently high that the material is in a condensed phase,  the energy shifts $\Delta E$  as a function of the density, $n$, 
have been very precisely determined through inelastic X-ray scattering experiments\cite{arms01,arms05}. Since the temperatures were also measured in these experiments, the corresponding molar volumes and hence densities, $n$,  were determined from a previously reported entirely experimentally derived equation of state \cite{glassford66}. It should be stressed that the entire pressure, temperature, volume and density analysis was entirely self-contained not invoking the measured values of $\Delta E$ or any unwarranted theoretical assumptions. Furthermore, it has been shown\cite{pyper21,pyper22} that non-empirical electronic structure computations accurately predict the experimentally determined density dependence the energy shifts for both the high densities studied in the  Arms {\it et al.} papers\cite{arms01,arms05} and for the lower density of helium under its own vapour pressure\cite{surko69}.
Each shift was calculated by adding non-empirically determined contributions from dispersion and short-range electron correlation to values predicted from \ab Hartree--Fock computations. The excellent agreement between the entirely independent experimental and theoretical studies shows that one has a fully trustworthy set of data in the bulk material against which the pore measurements can be compared.

\subsection{Relating pore, bubble and bulk behaviour from both experiment and theory}
\subsubsection{Methodology}
 The method for determining the relationship between the properties of a bubble or pore with those of the bulk ultimately relies on the following reasoning.
For bulk helium, the density $n$ producing any given shift $\Delta E$ is known. It therefore follows that the density in any individual bubble or pore having the same observed energy shift $\Delta E$ would have the same density $n$ provided that the behaviour of the helium in the bubble or pore was the same as that in the bulk.
 To determine whether the bulk or pore behaviour is actually the same as that of the bulk, the quantity $\Sigma\sr{theory}(n, E_0)$ is calculated from (\ref{eqn:Sigma_theory}) and compared with the experimental value $\Sigma\sr{exp}(E_0)$ (\ref{eqn:Sigma_exp}) for the same $\Delta E$. Equality of these two values would show that the behaviour of the helium in the pore or bubble is essentially the same as that of the bulk.
 
 The STEM measurements are sufficiently challenging as to cause the STEM data resulting from measurements on a series of bubbles or pores to be subject to considerable uncertainties.
Consequently, the bubble/pore results are most conveniently compared with those of the bulk by plotting the values of $\Delta E$ against $\Sigma\sr{exp}(E_0)$. If the helium behaviour in the bulk or pore was the same as that of the bulk, the plot of $\Delta E$ against the experimentally determined $\Sigma\sr{exp}(E_0)$ for the bubble/pore would coincide with that of $\Delta E$ against $\Sigma\sr{theory}(n, E_0)$ with both quantities computed for the bulk from the electronic structure calculations.

\subsubsection{Results of pore, bubbles and bulk comparisons}
\begin{figure}
\includegraphics[scale=0.5]{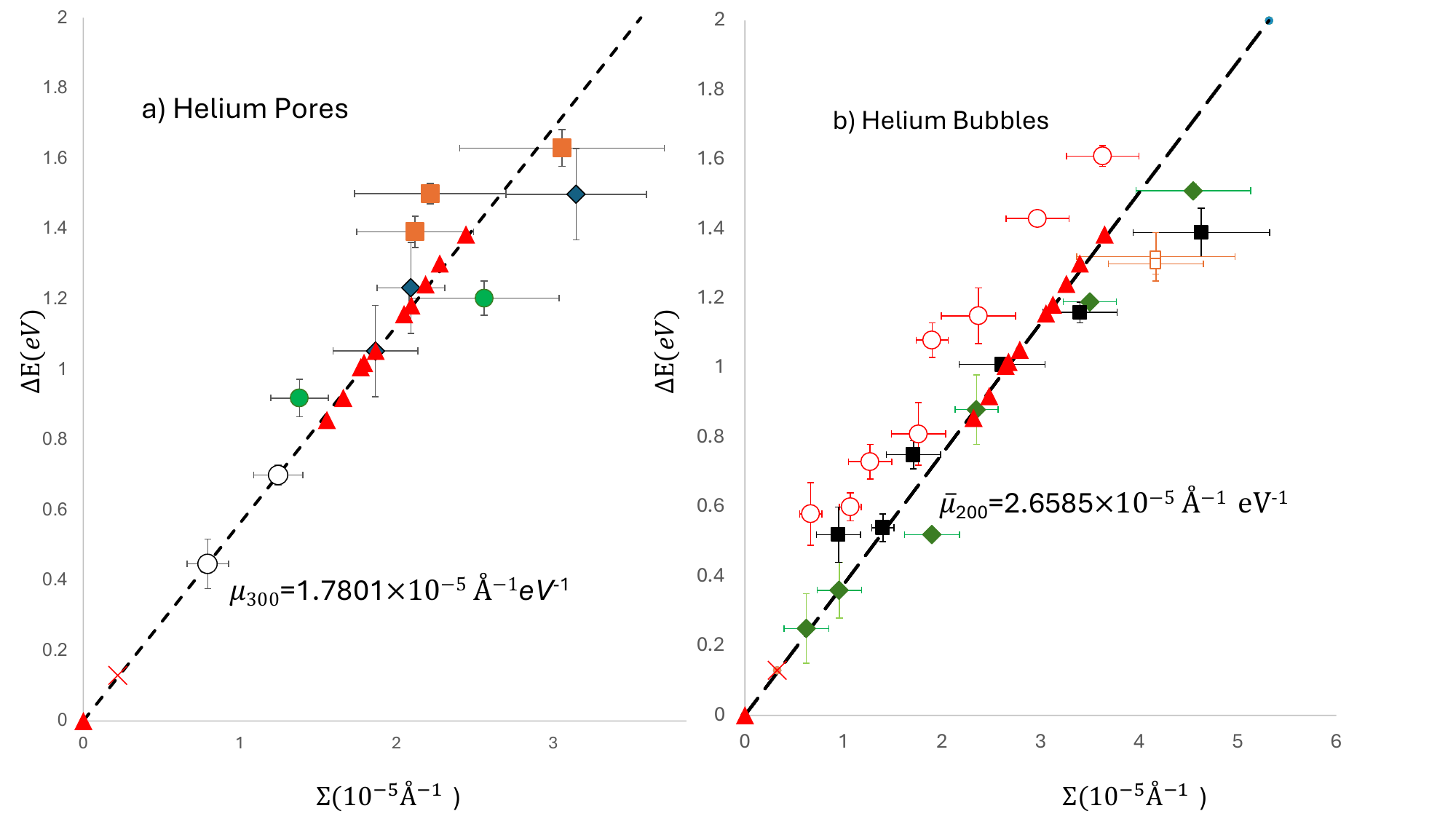}
\caption{\it{
Left panel: data from different pores coloured as in Figure \ref{systematic}; theoretical reference point: red cross at (0.22$\times10^{-5}$\colmang$^{-1}$, 0.13\,eV); linear fit to pore data (through the origin and the two points with lowest $\Delta E$), $\Sigma\sr{fit}(300\,\keV)=\mu_{300} \Delta E$ with slope $\mu_{300}=1.7801\times 10^{-5} \colmang^{-1} \text{eV}^{-1}$ :  black dashed line.\\
Right panel: Bubble measurements\cite{david22,pyper23} at 200 keV as follows: Bubble 1: red open circle;
bubble 2: green solid diamond; bubble 3: black solid rectangle; bubble 4 orange open rectangle; reference point\cite{pyper23} (0.32$\times10^{-5}$\colmang$^{-1}$, 0.13\,eV): red cross. The black dashed straight line through origin is given by $\bar{\Sigma}\sr{fit}(200\,\keV)=\bar{\mu}_{200} \Delta E$ with $\bar{\mu}_{200}=2.6585\times 10^{-5} \colmang^{-1} \text{eV}^{-1}$ from Equation \ref{eqn:barmu200_value}}.\\
Both panels: Computed results for the extended-fcc\cite{pyper22} description of the bulk: red filled triangles. }
\label{linearfits}
\end{figure}

 For the ten pore portions considered in this paper, the plot of $\Delta E$ against $\Sigma\sr{exp}(300\,\keV)$ shown in Figure \ref{linearfits}(a) agrees to within experimental error with the $\Sigma\sr{fit}(300\,\keV)$ predicted from the linear relation (\ref{linearNoArms}) derived by considering only the two pore portions having the smallest energy shifts. Furthermore, this liner pore relation passes essentially exactly though the plot of $\Delta E$ against the values $\Sigma\sr{theory}(300\,\keV)$ computed in bulk helium using (\ref{eqn:Sigma_theory}). This shows that the behaviour of the helium in the pores is essentially the same as that of the bulk.  
 The availability of results for both the pores and the previously studied bubbles enables the helium behaviours in these two systems to be compared directly using a new method without invoking any results for bulk helium.  

A  direct comparison of the bubble and pore results is complicated by the different beam energies in the  two microscope experiments. For the former $E_0=200$ keV while for the latter it was 300 keV.  
The plot in Figure \ref{linearfits}(a) shows that the experiential results for the pore data recorded at 300\,keV are well described by the linear relation, equation (\ref{linearNoArms}). This suggests that the $\Sigma\sr{exp}(200\,\keV)$ values that would have been measured for the pores using an impact energy of 200\,keV might be well described by a similar linear relation of the form
\begin{equation}
\bar{\Sigma}\sr{fit}(200\,\keV)=\bar\mu_{200}\Delta E,\label{eqn:barmu200}
\end{equation}
where the bars denote that $\bar\mu_{200}$ is to be derived from the directly experimentally determined $\mu_{300}$ for the pores. 
Following the same procedure used to determine $\mu_{300}$ for the pores, the quantity $\mu_{200}$, introduced later, was derived\cite{pyper23} for the bubbles by finding the straight line passing through the origin and the two points of lowest $\Delta E$.
Since $\Delta E$ is independent of $E_0$, the equations (\ref{linearNoArms}) and (\ref{eqn:barmu200}) can be unified into the form

\begin{equation}
\Delta E = \frac{\mu_{300}}{\Sigma\sr{fit}(300\,\keV)} =  \frac{\bar\mu_{200}}{\bar\Sigma\sr{fit}(200\,\keV)},\label{eqn:E_mus} 
\end{equation}
so
\begin{equation}
\bar\mu_{200} = \frac{\Sigma\sr{theory}(n, 200\,\keV)}{\Sigma\sr{theory}(n, 300\,\keV)}\mu_{300}=\frac{\sigma^{2p}(n, 200\,\keV)}{\sigma^{2p}(n, 300\,\keV)}\mu_{300},\label{eqn:barmu200_calc}
\end{equation}
where the explicit factors of $n$ in the definition (\ref{eqn:Sigma_theory}) of the $\Sigma\sr{theory}(n, E_0)$ cancel. It has already been established that the relation (\ref{colm11}) is valid for both the pore experiments at 300\,keV and the bubbles measurements at 200\,keV thus justifying the replacements of the $\Sigma\sr{fit}(E_0)$ by $\Sigma\sr{theory}(n, E_0)$ on passing from (\ref{eqn:E_mus}) to (\ref{eqn:barmu200_calc}).
 The computed values of $\sigma^{2p}(n, 200\,\keV)$ and $\sigma^{2p}(n, 300\,\keV)$ presented in Table I show that their ratios vary only slightly with density having an average value of 1.4935.
 This is very close to the value of $\frac32$ that would be predicted from (\ref{born1}) if the explicit factor of $E_0$ was the only impact-energy-dependent term. Introducing the computed ratio into (\ref{eqn:barmu200_calc}) shows that one has
\begin{equation}
\bar{\mu}_{200}=2.6585\times 10^{-5} \colmang^{-1} \text{eV}^{-1}.\label{eqn:barmu200_value}
\end{equation}%
These considerations show if the bubble and the pore properties are compatible we would expect
$\bar{\Sigma}\sr{fit}(200\,\keV)=\bar\mu_{200}\Delta E$ to be a good fit to the bubble  results as well as the reference corresponding to $n=0.0083$\dena, $\Delta E=0.13$ eV, $E_0=200$ keV, where the cross-section can be approximated to the free-atom value\cite{ral08} $\sigma^{2p}(0, \text{200  keV})=0.383 \times 10^{-3} \colmang^2$.
In  Figure \ref{linearfits}b  a comparison is shown for both the predictions for the bulk and results\cite{pyper23} for the bubbles with the linear approximation defined by the value of
$\bar{\mu}_{200}$ from (\ref{eqn:barmu200_value}).
These results strongly  suggest that the bubble and pore results are compatible with the only difference between the two sets of measurements being solely attributable to the different energies in the microscopes.
Our calculations and the comparison of the three environments strongly suggest  that the energy shift depends purely on the density of the aggregate.

Further evidence for the validity of the approach of introducing the $\bar\mu_{200}$ to investigate the helium properties is provided by comparing the $\bar\mu_{200}$ with that $\mu_{200}$ determined directly from the bubble data\cite{pyper23}.  This latter value was derived by fitting the $\Sigma\sr{exp}(200\,\keV)$ to the two points with the lowest value of $\Delta E$ after correction to ensure that the $\Delta E$ v. $\Sigma\sr{exp}(200\,\keV)$ line passes through the origin. The difference of a mere 2\% between the resulting $\mu_{200}$ value of $2.6074\times 10^{-5} \colmang^{-1} \text{eV}^{-1}$ and that just derived as $\bar\mu_{200}$ provides further evidence of the extremely close behaviours of helium in the bubbles and pores.

In  the theory\cite{pyper22} used here  it was assumed that the atom which was excited was at the centre of
of  a group of helium atoms arranged in a  fcc structured lattice and that  this group forms a lattice portion entirely surrounded by the other helium atoms. It was implicitly assumed  that the lattice portion of interest was sufficiently far from the boundary of the entire cluster that edge effects could  be neglected.
Naively one would expect the \quotes{top} pore segments (\ie the \quotes{tail of the tadpole}) to have a higher percentage of pixels close to the encapsulating silicon compared
to the \quotes{bottom} pore segment (the \quotes{head of the tadpole}).
The fact that this theory works so well for all the pore portions seems to suggest that if edge-effects exist they will be confined to a very narrow region close to the encapsulating material.
The results of this section show that the observed energy shifts are directly related to the density in the same way in bubble, pore and bulk. Further, the  success of the
linear approximation $\bar{\Sigma}\sr{fit}(E_0)$ suggests that we have an effective way of estimating  the key characteristics of the pores. This will be explored further below.
\section{Characteristic Properties of the Pores}
\subsection{Pore Depths}
\subsubsection{Standard depth predictions}
Values of pores depths are not only of direct interest for applications in the manufacturing of thin films but are also needed in the present investigation to show that the behaviour of the helium in the pores is essentially the same as that of the bulk. The interest for applications motivates elucidating the errors in the pore depths derived from the approach, known as the log ratio method\cite{egerton11}, described in section \ref{sec:det_h}. The result, that these errors can be quite appreciable, provides a strong motivation for developing an alternative method for deducing pore depths. 
In the approach described in section \ref{sec:det_h}, each pore depth, $h\sr{exp}$ is evaluated as the average of the predictions produced by using (\ref{eqn:h_calc}) based on the log ratio method.
The resulting predictions for the $h\sr{exp}$, presented in the seventh numerical column in Table \ref{tbl:h_exp},
 contain uncertainties arising from both experimental errors in the intensity and pore depth in $\ln$ term in (\ref{eqn:h_calc}) as well as those arising from the estimated\cite{malis88} 20\% error in the silicon mean free path $\lambda\sr{m}$. Standard analysis yields, for each $h\sr{exp}$, the error $\delta h\sr{expI}$ that would arise if the uncertainties in the value of $\lambda\sr{m}$ were not considered. The absolute values of the resulting errors, $\delta h\sr{expI}$, ranging from 6.3\% to 19\%, are assembled in the eighth numerical column of Table \ref{tbl:h_exp}. Consideration of the 20\% error in $\lambda\sr{m}$ in conjunction with those arising in the $\ln$ term enables the overall fractional errors in the $h\sr{exp}$ to be derived as the square root of the sums of the squares of the individual fractional errors, via
\begin{equation}
\frac{\delta h\sr{exp}}{h\sr{exp}} = \sqrt{\left(\frac{\delta h\sr{expI}}{h\sr{exp}}\right)^2 + (0.2)^2}\label{eqn:dhexp}
\end{equation}
The absolute errors, $\delta h\sr{exp}$, presented in the ninth numerical column in Table \ref{tbl:h_exp} are derived from the fractional errors predicted from (\ref{eqn:dhexp}), which range from 21\% to 28\%.

\begin{table}[h!]
\caption{Two derivations of pores depth values and errors}
\begin{tabular}{|c|c|c|c|c|c|c|c|c|c|c|c|}
\hline
Pore & Portion & $I\sr{He}$ & $\delta I\sr{He}$ & $\Delta E$ & $\delta\Delta E$ & $\bar h$ & $\delta \bar h$ & $h\sr{exp}$ & $\delta h\sr{expI}$ & $\delta h\sr{exp}$\\
     &         &       (a.u.) &   (a.u.)            &  (eV)      &       (eV)           &           (\AA) &    (\AA)   &          (\AA)     &         (\AA) & (\AA)\\
\hline

1& Top  &  59638 &    6419 &  1.4979 & 0.0483 & 35.374 & 3.975 &  29.969 & 2.8112 &  6.6202  \\\hline
 &Middle&  78365 &    5555 &  1.2319 & 0.0844 & 54.401 & 5.364 &  57.045 & 4.2734 &  12.183  \\\hline
 &Bottom&  68395 &    8897 &  1.0521 & 0.0521 & 55.732 & 7.757 &  55.944 & 3.5548 &  11.740  \\\hline
2&Top   &  99923 &   10442 &  1.6292 & 0.0522 & 44.162 & 4.827 &  41.895 & 7.7945 &  11.444  \\\hline
 &Middle& 156723 &   14908 &  1.4996 & 0.0287 & 70.085 & 6.800 &  84.440 & 16.548 &  23.644  \\\hline
 &Bottom& 147678 &   14446 &  1.3912 & 0.0444 & 71.153 & 7.321 &  83.154 & 12.202 &  20.627  \\\hline
3&Top   &  67750 &    7149 &  0.6993 & 0.0285 & 58.183 & 6.359 &  58.180 & 4.0092 & 12.290  \\\hline
 &Bottom&  45418 &    5781 &  0.4466 & 0.0704 & 60.893 &12.335 &  60.906 & 6.5392 & 13.825 \\\hline 
4&Top   &  92617 &   10139 &  1.2024 & 0.0486 & 56.142 & 6.551 &  46.945 & 7.0976 & 11.770 \\\hline 
 &Bottom&  73197 &    8987 &  0.9184 & 0.0537 & 56.274 & 7.653 &  66.615 & 3.4820 & 13.770 \\\hline 
\end{tabular}
 \label{tbl:h_exp}
\end{table}

\subsubsection{A new method for determining pore depths}
In the determination of $\Sigma\sr{exp}(E_0)$ through its definition (\ref{eqn:Sigma_exp}), the large errors in the pore depths predicted from (\ref{eqn:h_calc}) are combined with those arising from the experimental measurements of $I\sr{He}$. The fractional errors in the latter, ranging from 7\% to 12\%, are derived by dividing their absolute errors $\delta I\sr{He}$ presented in the second numerical column in Table \ref{tbl:h_exp} by the $I\sr{He}$ in the first column. The errors in the resulting values of $\Sigma\sr{exp}(n,E_0)$ cause the significant scatter in this quantity, as manifested in Figures \ref{systematic} and \ref{linearfits}. However, the strong evidence presented in section \ref{sec:pores_etc}, that the helium behaviour in the bulk is the same as that in both the bubbles and pores coupled with the ability to predict the bubble behaviour directly from that of the pores via linear relations of the type (\ref{linearNoArms}) and (\ref{eqn:barmu200}), justifies the alternative and new method, presented below, for determining pore portion depths. 
This method proceeds by equating, for each pore portion, the $\Sigma\sr{exp}(n,E_0)$ result derived from (\ref{eqn:Sigma_exp}) to that predicted from the $\Sigma\sr{fit}(n,E_0)$ description using the experimentally measured value of $\Delta E$. This procedure, followed by rearranging the resultant equation yields the new prediction, denoted $\bar{h}$, for the pore portion depth.
\begin{equation}
\frac{I\sr{He}}{I\sr{ZLP}\bar h}=\mu_{300}\Delta E,
\end{equation}
so
\begin{equation}
\bar{h}=\frac{I\sr{He}}{I\sr{ZLP}\mu_{300}\Delta E}.\label{eqn:barh}
\end{equation}
Here the beam energy of 300\,keV in the pore experiments has been introduced.

Comparison of the predictions derived from (\ref{eqn:h_calc}) with those from (\ref{eqn:barh}) shows the errors 6\% to 19\% arising from the $\ln$ term in the former are absent but are replaced by the 7\% to 13\% errors in $I\sr{He}$ on using (\ref{eqn:barh}). The much more significant differences between the total errors in the pore portion depths predicted by using (\ref{eqn:barh}) rather than (\ref{eqn:h_calc}) arise from the absence of 20\% errors in $\lambda\sr{m}$ being replaced by errors $\delta\Delta E$ in $\Delta E$. The third and fourth numerical columns in table \ref{tbl:h_exp} show that the fractional errors in the $\Delta E$ are much smaller, lying between just 3\% and 7\%,  except for the 16\% error for the bottom of pore 3. The total errors in the predictions for the pore portion depths are given by
\begin{equation}
\frac{\delta h}{h}=\sqrt{\left(\frac{\delta I\sr{He}}{I\sr{He}}\right)^2+\left(\frac{\delta\Delta E}{\Delta E}\right)^2}.\label{eqn:errh}
\end{equation}

The values of $I\sr{ZLP}$ and  $\delta I\sr{ZLP}$ presented in the supplementary material show that the fractional errors in the $I\sr{ZLP}$ are no more than 1\% and can therefore be neglected in this error analysis. Although the errors in the pore depths measured by the log ratio method are appreciable, they yield values of $\Sigma\sr{exp}(E_0)$ to an accuracy sufficient to conclude that the helium behaviour in a pore is the same as that in the bulk. Once this has been established, the $h\sr{exp}$ values can be discarded and replaced by using (\ref{eqn:barh}) to derive the pore portion depths more accurately. 

The pore portion depths predicted using (\ref{eqn:barh}) are assembled in the fifth numerical column in Table \ref{tbl:h_exp} with their total errors evaluated from (\ref{eqn:errh}) appearing in the sixth column. For all the pore portions, excepting that at the bottom of pore 3, the fractional errors lying between 9\% and 14\% are much less than those over 20\% arising in the log ratio method. Even for the bottom of pore 3, the 20.2\% error in the prediction for $h$ does not exceed the 22.7\% error produced by the log ratio method. The above error analysis shows that it is this new method which produces the more accurate values for the depths of the pore portions.
Given sufficiently accurate experimental data, the method outlined here could be employed to determine the volume, by measuring the depth and surface area of each individual pixel and summing the resulting pixel volumes.
\subsection{Densities and Pressures}
While this paper is primarily concerned with the depth and density of each pore, there are situations where pressure can be important. For example, for studies involving  applications in magnetism and catalysis, lattice-strain effects in the encapsulating material may be significant.\cite{giarratano18}.

The result shown in section \ref{sec:pores_etc} that the behaviour of the helium in a pore is the same as that of the bulk has the consequence that the energy shift is a unique function of the density. This function was derived\cite{pyper22} by demanding that it reproduced the experimental results for both bulk liquid helium under its own vapour pressure\cite{surko69} and for the bulk condensed phases under higher pressures\cite{arms01,arms05} as well as passing through the origin. The resulting relation yielding $n$ in atoms $\colmang^{-3}$ given $\Delta E$ in eV is
\begin{equation}
n=0.044 (\Delta E)^{0.72},\label{eqn:colm_n}
\end{equation}
from which the densities, see Table \ref{pressure1} can be predicted from the experimental values of the energy shifts presented in Table \ref{tbl:h_exp} for the pore portions. Six out of the ten densities and energy shifts thus derived lie within the range of those measured by Arms \textit{et al.}\cite{arms01,arms05} for the bulk condensed phases at higher pressures. This ensures that the densities will be accurately predicted from (\ref{eqn:colm_n}). Furthermore, the densities and shifts predicted for the two most dense pores lie only very slightly outside that range studied in [\!\!\citenum{arms01},\!\!\citenum{arms05}] whilst the lowest shift and density  values derived for the least dense pore lie close to the experimentally determined\cite{surko69} values for the liquid helium under its own vapour pressure. 

 For helium at the 300\,K temperature of the pore measurements, the pressures are accurately predicted from the equation of state presented by Mills \textit{et al.}\cite{mills80}. The was derived by parameterizing experimental values measured for pressures ranging from 200\,MPa to 2000\,MPa.
For nine of out the ten pore portions, the pressures will be very accurately predicted because these lie within this range studied in [\!\!\citenum{mills80}]. Furthermore, the lowest pressure predicted for any pore portion lies only slightly below this range. 
 The degassing effects of the probing electron beam will be of least significance for the top portion of each pore because this beam is initially focused on that region.
The difficulties discussed in section \ref{exp_motiv}, with the previously reported densities, show that both these, as well as the pressures derived from them, have been superseded by the results assembled in Table \ref{pressure1}.

\FloatBarrier
\begin{table}[h!]
\caption{ Calculated density and pressure for the pore portions. Densities were calculated using (\ref{eqn:colm_n}), and pressures using the Mills equation\cite{mills80} at 300\,K.\label{pressure1}}
\begin{tabular}{|c|c|c|c|c|c|c|c|c|c|c|}
\hline
 Pore &1&1&1&2&2&2&3&3&4&4\\
Portion&top&middle&bottom&top&middle&bottom&top&bottom&top&bottom\\\hline
Density, $n$ (atoms/$\colmang^3$) & 0.05886 & 0.05133 & 0.04564 & 0.06253 & 0.05890 & 0.05581 & 0.03401 & 0.02463 & 0.05025 & 0.04138 \\\hline
Pressure (MPa) & 978.1 & 707.9 & 546.9 & 1137.9 & 980.0 & 860.0 & 307.3 & 179.1 & 674.7 & 446.7\\\hline
 \end{tabular}
 \end{table}
 \FloatBarrier

It is clear that, at any given time, the helium in all the different portions of a single pore must exert the same pressure. However, the pore becomes progressively degassed by the probing beam as each portion is examined sequentially in time. The significance of the degassing is seen by comparing the densities and pressures presented in Table \ref{pressure1} for the top pore portions with those for the middle and bottom portions.

\section{Conclusions and Summary}
A conjunction of theory and STEM/EELS experiments has been used to investigate both the behaviour of helium encapsulated in nano pores in amorphous silicon as well the properties of the nanopores themselves. This study extends to these pores, the previous investigations\cite{david11,david14,alix15,alix18,pyper23} of helium encapsulated as spherical bubbles in silicon. 

It has been shown in this present paper that, in the non-spherical pores of complex shapes, the behaviour of the helium is the same as that already established for the spherical bubbles\cite{pyper22} with both behaviours being the same as that for the bulk condensed phases both under high pressures and for the liquid in equilibrium with its vapour. The theory used\cite{pyper01,pyper21}, which is entirely non empirical containing no adjustable parameters, is validated by the excellent agreement of its predictions\cite{pyper22} with the fully independently and experimentally determined results\cite{surko69,arms01,arms05} for bulk helium. 

The demonstration that the helium behaviour in the bubbles and pores is the same as that of the bulk required experimental measurement of both the 1s$^2\rightarrow$1s2p($^1$P) excitation energy as well as of another quantity constructed from experimentally measured intensities of the scattered electron beams. The result that these two properties are shown to be linearly related underlies two further and new developments. Firstly, it has been shown the difference between measurements from microscopes working at different electron impact energies could be simply related by the ratios of the helium cross sections.
 This allows the behaviour of the helium encapsulated in different materials to be compared directly without recourse to data on bulk helium.
 In this comparison the helium behaviours in the non-spherical pores and the spherical bubbles were shown to be identical, thus reinforcing the same conclusion reached through the comparison of the individual behaviours with that of the bulk.
 Secondly, having established that the helium behaviour in the pores was the same as that in the bulk, a new linear relationship  between the energy shifts and the  $\Sigma$, circumventing the significant scatter in the individual  experimental values  of the latter,
 was used to derive an expression for depths in the pores. The individual $\Sigma_{exp}$ 
  have much larger errors than  those in the environmentally induced enhancements of the 1s$^2\rightarrow$1s2p($^1$P) excitation energy whose values enter the new pore depth expression. This was used to present numerical values the depths of the pore portions more accurately than the standard log ratio method. This new method of measuring depths should also be applicable to faceted bubbles.
Once it is established that the helium behaviour in the pores or bubbles is the same as that of the bulk, the helium density is uniquely determined by the energy shift.  This was used to determine the density in each of the ten pore sections studied from which the helium pressures were deduced.

\section{Acknowledgements}
 This research was funded by: (i) MICIN/AEI/10.13039/501100011033, FEDER-EU, project PID2021-124439NB-I00; (ii) EMERGIA 2021 program from the Andalusian Regional Government --- Junta de Andalucía (EMC21\_00427 grant to B.L.). C.T.W. is grateful for support from the Royal Society(IES\textbackslash R1\textbackslash 231215).
\section*{Data Avalability}
All the data needed to reproduce the results presented in this paper is contained in the Supplementary Material.

\bibliographystyle{unsrtnat}
\bibliography{confined_mat.bib}{}
\end{document}